\documentclass{iopart}

\usepackage{graphicx}% Include figure files
\usepackage{dcolumn}% Align table columns on decimal point
\usepackage{bm}% bold math
\usepackage{iopams}

\expandafter\let\csname equation*\endcsname\relax
\expandafter\let\csname endequation*\endcsname\relax
\usepackage{amsmath,amssymb}

\usepackage{color}% 
\usepackage{upgreek,soul}
\newcommand{\lb}{\left(}
\newcommand{\rb}{\right)}
\newcommand{\lsb}{\left[}
\newcommand{\rsb}{\right]}
\newcommand{\half}{\frac12}
\newcommand{\const}{\mbox{const}}
\newcommand{\bmr}{\mathbf{r}}
\newcommand{\bk}{\mathbf{k}}
\newcommand{\bmxi}{\bm{\xi}}
\newcommand{\bJ}{\mathbf{J}}
\newcommand{\bV}{\mathbf{V}}
\newcommand{\bv}{\mathbf{v}}
\newcommand{\DD}{\Delta^2}
\newcommand{\brt}{\mathbf{r_t}}
\newcommand{\pdz}{\partial_z}
\newcommand{\mum}{\upmu\text{m}}
\newcommand{\mums}{\upmu\text{m/s}}

\usepackage[latin1]{inputenc}
\begin{document}

\title{Stirring by swimmers in confined microenvironments}

\author{Dmitri O. Pushkin and Julia M. Yeomans}
\address{The Rudolf Peierls Centre for Theoretical Physics, 1 Keble Road, Oxford, OX1 3NP, UK}
\ead{mitya.pushkin@physics.ox.ac.uk}

\date{\today}

\begin{abstract}
We consider the tracer diffusion $D_{rr}$ that arises from the run-and-tumble motion of low Reynolds number swimmers, such as bacteria. In unbounded dilute suspensions, where the dipole swimmers move in uncorrelated runs of length $\lambda$, an exact solution showed that $D_{rr}$ is independent of $\lambda$. Here we verify this result in numerical simulations for a particular model swimmer, the spherical squirmer. We also note that in confined microenvironments, such as microscopic droplets, microfluidic devices and bacterial microzones in marine ecosystems, the size of the system can be comparable to $\lambda$. We show that this effect alone reduces the value of $D_{rr}$ in comparison to its bulk value, and predict a scaling form for its relative decrease.
\end{abstract}

\maketitle

\section{Introduction \label{s:Intro}}

As microswimmers, such as bacteria, algae or active colloids, move they produce long-range velocity fields which stir the surrounding fluid. As a result particles and biofilaments suspended in the fluid diffuse more quickly, thus helping to ensure an enhanced nutrient supply.  Following the early studies of mixing in concentrated microswimmer suspensions \cite{Wu00,Dombrowski04,Sokolov09}, recent experiments have demonstrated enhanced tracer diffusion in dilute suspensions of {\em Chlamydomonas reinhardtii}, {\em Escherichia coli} and self-propelled particles \cite{Leptos,Kurtuldu,Mino11,Mino12,Poon13}. Simulations 
have found similar behaviour \cite{Underhill08,IshiPedley10,LinChildress11} and microfluidic devices exploiting the enhanced transport due to motile organisms have been suggested \cite{KimBreuer04,Kimbreuer07}. However, theoretical description of fluctuations and tracer mixing in active systems remains a challenge even for very dilute suspensions of microswimmers.

The Reynolds number associated with bacterial swimming is $\sim 10^{-4}-10^{-6}$. Therefore the flow fields that result from the motion obey the Stokes equations and the far velocity field can be described by a multipole expansion. The leading order term in this expansion, the Stokeslet (or Oseen tensor), which decays with distance $\sim r^{-1}$, is the flow field resulting from a point force acting on the fluid. However biological swimmers, which are usually sufficiently small that gravity can be neglected, move autonomously and therefore have no resultant force or torque acting upon them. Hence the Stokeslet term is zero and the flow field produced by the microswimmers contains only higher order multipoles, for example dipolar contributions, $\sim 1/r^2$, and quadrupolar terms, $\sim 1/r^3$.

\begin{figure}
\centerline{\includegraphics[width=.6\columnwidth]{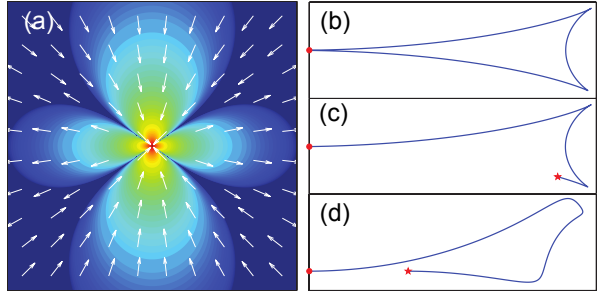}}
\caption{ (a) The angular dependence of the dipolar flow field. The velocity decays as $r^{-2}$ where $r$ is the distance from the swimmer. (b) A typical, closed-loop tracer trajectory for an infinite, straight swimmer path and tracer velocity $\ll$ swimmer velocity. (c) A typical trajectory for a finite swimmer path (d) A typical entrained trajectory, for an infinite swimmer path, and tracer close to the swimmer \cite{Pushkin13a}.
}
\label{f:tracer_paths}
\end{figure}

The absence of the Stokeslet term has important repercussions for the way in which tracer particles are advected by swimmers. The angular dependences of the dipolar velocity field -- shown in Fig.~\ref{f:tracer_paths} -- and of higher order multipoles of the flow field lead to loop-like tracer trajectories. For a distant swimmer, moving along an infinite straight trajectory these loops are closed \cite{Dunkel10,Pushkin13a}.

The paths of bacteria or active colloids are, however, far from infinite straight lines. For example, periodic tumbling (abrupt and substantial changes in direction) is a well established mechanism by which microorganisms such as {\it E. coli} can move preferentially  along chemical gradients. Even in the absence of tumbling, microswimmers typically have curved paths due to rotational diffusion or non-symmetric swimming strokes. For non-infinite swimmer trajectories tracers no longer move in closed loops and the swimmer reorientations cause enhanced diffusion. In our recent work \cite{PushPRL}, building on a model suggested in \cite{LinChildress11}, we found an exact analytical expression for the tracer diffusion coefficient due to random reorientations $D_{rr}$ of dipole swimmers in an unbounded suspension of the swimmers. We explained that the somewhat paradoxical independence of $D_{rr}$ on the random reorientation length $\lambda$ is a consequence of a fortuitous balance between the dimensionality of space $d=3$ and the character of the hydrodynamic flow field around a force-free dipole swimmer. By contrast, other than the second moments of the tracer distribution, and thus the distribution itself, will depend on $\lambda$. Also, the conclusion that $D_{rr}$ is independent of $\lambda$ holds only for $ \lambda \ll L$, where $L$ is the characteristic size of the system. It remains to be explained what happens to the tracer diffusion when the trajectories' persistence length becomes comparable to the system size. This question also has a practical merit: in many technological and environmental circumstances the swimmers are confined to microenvironments, such as microscopic droplets, microfluidic devices \cite{Stoker_microfluidics}, microscopic pores in soft agar used for biological assays \cite{Poon_agar}, and nutrient-rich microzones, which play prominent role in marine microbial ecosystems \cite{MarineEcosystems07,MarineEcosystems85}.   

The current paper is devoted to elucidating these questions. In section \ref{s:Stirring} we review stirring mechanisms and the analytical theory of tracer diffusion due to random reorientations. In section \ref{s:Statistics} we give a theoretical analysis of the form of the tracer displacement distribution for finite times.  Section \ref{s:Numerics} is devoted to comparing our predictions to the results of numerical simulations. In particular, we discuss the effects of the finite system size on $D_{rr}$. Our main conclusions are discussed in section \ref{s:Discussion}.  

%---------
\section{Stirring mechanisms \label{s:Stirring}}

\subsection{Tracer diffusion coefficient for dilute suspensions of swimmers}

Before discussing details of stirring mechanisms, we comment on the meaning of the limit of very dilute swimmer suspensions. Essentially, this limit guarantees that interactions between microswimmers are negligible so that their effects on tracer transport are additive. Then the effective tracer diffusion coefficient must be proportional to the concentration of swimmers $n$. In addition, dimensional considerations require that the effective diffusion coefficient is proportional to the swimmer velocity. Therefore, $D \propto n V$. It was suggested in \cite{Mino11} to call the latter quantity {\it active flux}. Linear scaling with active flux has been confirmed in experiments \cite{Leptos,Mino11,Mino12,Poon13} and simulations \cite{LinChildress11}. It was also derived in \cite{Underhill08} using the Green--Kubo formalism and in \cite{LinChildress11} using a random walk reasoning. In these cases the characteristic volume fraction of swimmers $\Phi \sim 10^{-6}-10^{-3}$. (Note that in most abundant bacterial marine environments $\Phi$ usually does not exceed $10^{-6}$ \cite{MarineEcosystems07}.)

However, one should bear in mind that due to the long-range nature of hydrodynamic interactions, the correlations between swimmers may build up quickly for seemingly low swimmer concentrations. The notable difference in the diffusion coefficient between suspensions of pushers and pullers which is due to interactions between swimmers is already apparent for $\Phi \sim 0.1$ \cite{Underhill11}. The increasing role of the correlations at higher $\Phi$ will be reflected as a nonlinear dependence of $D$ on $n$: $D \propto V (n + b_2 \, n^2)$, where $b_2$ is a constant.

\subsection{Velocity fields of dipole swimmers}

For dipolar swimmers with velocity $\bV=V \bk, \, \| \bk \|=1$ the leading order term in the far-field expansion of the velocity is the stresslet
\begin{equation}
\bv (\bmr)\approx - \kappa \, \bk \cdot (\bk \cdot \nabla)  \bJ(\bmr),
\label{eq:stresslet}
\end{equation}
 where $\bmr$ is the radius-vector with the origin at the swimmer, $\kappa$ is the swimmer dipole strength, and
$\bJ(\bmr)$ is the Oseen tensor,
\begin{equation}
\bJ(\bmr)=\frac{\mathbf{I}}{r}+\frac{\bmr \bmr}{r^3}.
\end{equation}
(In this notation the fluid viscosity and numerical constants are adsorbed in $\kappa$.)  The swimmer velocity fields are extensile for $\kappa >0$ (e.g. for `pushers' such as {\em E. Coli}) and contractile for $\kappa <0$ (e.g. for `pullers' such as {\em C. reinhardtii}).
The validity of the expression (\ref{eq:stresslet}) is based on the assumption that the resultants of the drag and the propulsive forces are parallel to the swimming direction \cite{Pushkin13a} as has been verified experimentally for several types of biological microswimmers \cite{Drescher10,Drescher11}.

%---------------
\subsection{Stirring due to entrainment}

The trajectories of tracers advected by distant swimmers moving along an infinite straight path are closed loops. The loops cease to be closed when the swimmers' trajectories are curved or tumble or when the swimmer happens to pass so close to a tracer that the latter gains a velocity comparable to that of the swimmer and, as a result, is {\it entrained} along or opposite to the swimming direction. While in a real suspension of swimmers both effects occur simultaneously, in dilute swimmers suspensions it turns out helpful to distinguish them as separate stirring mechanisms having distinct features and physical meaning.
 
The entrainment of a tracer by a swimmer is defined as the displacement of the tracer as the swimmer passes by on an infinite straight trajectory. If a fixed point were considered instead of the tracer, the fluid velocity induced at this point by the swimmer integrated over time would result in strictly zero displacement due to the symmetry of the swimmer velocity field \cite{Pushkin13a}. However, as the tracer is {\it entrained} by the swimmer, the balance of times it spends in front and behind the swimmer is broken and results in a finite tracer displacement along or opposite to the swimming direction. Mathematically, this effect arises due to the Lagrangian contribution to the tracer velocity field and is a relative of the Stokes drift phenomenon in time-periodic flows \cite{Eames99}. Physically, this effect can be conveniently quantified by considering the fluid volume displaced by the swimmer across a plane normal to the swimming direction, the so-called Darwin or Lagrangian drift \cite{Darwin}. While the latter turns out to be rather sensitive to the swimmer near-field, its magnitude remains of the order of the swimmer volume. Also, the characteristic entrainment length typically remains of the order of the swimmer size. These considerations allow an order of magnitude estimate of the effective tracer diffusion due to entrainment $D_{entr}$ in an uncorrelated uniform suspension of swimmers \cite{Pushkin13a}:
\begin{eqnarray}
D_{entr} \sim  \frac{2 \pi}{9} a^{4} \, n \, V .
\label{e:DEntrScaling}
\end{eqnarray} 

While the effective diffusion coefficient is the major physical parameter characterising mixing of an {\it ensemble} of tracers, the character of the random walk performed by an {\it individual} tracer via the entrainment mechanism merits additional comment: in dilute suspensions of swimmers the flights of the tracer due to entrainment by a swimmer are relatively rare events. Indeed, the effect of entrainment weakens rather quickly as the distance between the tracer and the swimmer grows. Let us estimate the mean time between passages of a swimmer within two swimmer diameters from a given tracer, $\tau_c$. The mean free path of a swimmer $l_c$ can be estimated from $16 \pi a^2 n l=1$. Hence,
\begin{eqnarray}
\tau_c =\frac{l_c}{V} \simeq \frac{a}{12 \Phi V} \gg a/V \quad 
\text{when} \quad \Phi \ll 1.
\label{e:tau_c}
\end{eqnarray}
For example, in the experiments with suspensions of {\it C. reinhardtii} \cite{Leptos}, $a \sim 5 \mum$, $V \sim 100 \mums$ and $\Phi \sim 10^{-3}-10^{-2}$. Hence, $\tau_c \sim 0.4-4$s. As the experiment duration was less than $0.3$s, we conclude that a relatively small (but non-negligible) fraction of the observed tracers could have been displaced due to entrainment by the passing swimmers during the experiment. On the other hand in the experiments with suspensions of {\it E. coli} \cite{Poon13}, $a \sim 1 \mum$, $V \sim 20 \mums$, $\Phi \sim 10^{-3}$ and hence $\tau_c \sim 0.4-4$s, while the duration of the experiments $\sim 10$s. Therefore most tracers must have experienced entrainment by bacteria in these experiments.

\subsection{Stirring due to the curvature of swimmer trajectories \label{s:reorient}}

The displacement of a tracer due to a finite persistence length of a swimmer trajectory is qualitatively different from the entrainment. Indeed, for a finite straight run of a swimmer, the tracer displacement $\Delta \brt$ can be calculated as
\begin{eqnarray}
\Delta \brt = \int_i^f \bv(\brt-\bk z(t)) dt.
\end{eqnarray}
Here $t$ is time, $z(t)$ is the swimmer coordinate along the swimming direction $\bk$ and the origin is assumed to lie on the line of swimming. In the far-field the swimmer velocity is given by(\ref{eq:stresslet}), $dz=V\,dt$, and the time variation of $\brt$ in the integrand, i.e. the tracer entrainment, can be neglected as the induced tracer velocity is much smaller than the swimmer velocity. Then,
\begin{eqnarray}
\Delta \brt = -\frac{\kappa}{V} \bk \cdot \int_i^f \pdz \bJ (\brt - \bk z) dz = -\frac{\kappa}{V} \bk \cdot \bJ (\bmr)  \bigg|_{\mathbf{r_i}}^{\mathbf{\bmr_f}},
\label{e:drtracer}
\end{eqnarray} 
where $\mathbf{r_i}$ and $\mathbf{r_f}$ are the initial and final distances between the swimmer and the tracer respectively. 

The effects of finite persistence length of swimmers trajectories on tracer mixing in a dilute suspension of dipolar swimmers were first assessed by Lin, Thiffeault and Childress \cite{LinChildress11}. They assumed a uniform and isotropic ensemble of uncorrelated swimmers moving in straight runs of length $\lambda$, followed by instantaneous random changes of swimming direction. The consecutive runs were assumed to be statistically independent. Using a combination of analytical and numerical methods the authors found that ``the largest contributions to tracer displacement, and hence to mixing, arise from random changes of direction of swimming and are dominated by the far-field stresslet term". This conclusion, in particular, provides {\it a posteriori} justification for considering only the dominant far-field (dipole) term in defining the mixing efficiency due to reorientations and leads to the scaling law for the effective diffusion coefficient
\begin{eqnarray}
D_{rr} \approx A \, \beta^2 \, a^4 \, n \, V.
\label{e:scalDrr}
\end{eqnarray}
Here the dimensionless swimmer dipole strength $\beta=\kappa/(V a^2)$ is based on the characteristic swimmer size $a$ and the value $A \approx 3.7$ was found in \cite{LinChildress11} by fitting the numerical results.

This important result at first appears paradoxical, as the tracer diffusion due to swimmer reorientations turns out to be independent of the reorientation length $\lambda$. Should it not vanish as $\lambda \to \infty$? Resolution of this paradox is one of the purposes of the current paper.

\subsection{Analytical theory of stirring due to reorientations}

In our previous work \cite{PushPRL} we revisited the random reorientations model of Lin, Thieffeault and Childress. In particular, we found an exact analytical solution of this model for a suspension of dipole swimmers in a domain of infinite size. Analytical treatment of contributions of different relative positions of the tracer and the swimmer path segments was made possible by using elliptic coordinates with the origin at the mid-point of the swimmer trajectory and the left and right foci at the initial and final positions of the swimmer, respectively:
\numparts
\begin{eqnarray}
&x& = a_1 \, \cosh \mu \, \cos  \nu = a_1 \, \sigma  \, \tau, \label{e:elliptica}\\
&y& = a_1 \, \sinh \mu \, \sin  \nu, \;
y^2 = a_1^2 \, (\sigma^2-1) \, (1-\tau^2) ,
\label{e:ellipticb}
\end{eqnarray}
\endnumparts
where $a_1=\lambda/2$, $\mu \ge 0$, $0 \le \nu \le 2 \, \pi$, $\sigma \ge 1$, and $-1 \le \tau \le 1$. In these coordinates, the tracer displacement (\ref{e:drtracer}) can be expressed as a rational function of  $\sigma$ and $\tau$:
\begin{eqnarray}
\DD(\sigma,\tau) &=& \Delta \mathbf{r_t} ^2=
\lb \frac{\kappa}{V} \rb^2 
\frac{4}{a_1^2 \, (\sigma^2-\tau^2)^4 } \, \nonumber\\
 && \lsb \lb 1-3 \, \tau^2 \rb ^2 \sigma^4 
   + \lb 3 \, \tau^2 -1 \rb   \lb \tau^2 +2 \tau -1 \rb   \right. \nonumber \\ 
 && \left. \lb \tau^2 -2 \tau -1  \rb   \sigma^2 
    + \tau^2 \lb  1 + \tau^2 \rb^2
\rsb .
\label{e:D3_elliptic}
\end{eqnarray}

To obtain the diffusion coefficient we assume a uniform isotropic distribution of swimmers that move in straight runs of length $\lambda$. We assume no interactions between the swimmers and statistical independence of the consecutive runs. Then the mean squared displacement of the tracer $\langle \DD \rangle$ can be obtained by summing over all possible straight segments of swimmer paths
\begin{equation}
\langle \DD \rangle \, = n_s \, \int d^3 \mathbf{r_i} \, \DD(\mathbf{r_i})
\label{e:D}
\end{equation}
where $n_s$ is the concentration of the segments and $\mathbf{r_i}$ is the initial distance between the tracer and the swimmer.
When the time $t$ is greater that the reorientation time, each swimmer contributes on average $Vt/\lambda$ segments. Therefore the diffusion coefficient $D_{rr}$ due to uncorrelated, random swimmer reorientations is
\begin{equation}
D_{rr}=\langle \DD \rangle /\,( 2 \, d \, t) \, =  \frac{1}{d} \frac{n \, V}{\lambda} \, \int d^3 \mathbf{r_i} \, \DD(\mathbf{r_i})
\label{e:DD}
\end{equation}
where $n$ is the number density of swimmers. 

Substituting in the expression (\ref{e:D3_elliptic}) for the displacement due to a single swimmer and transforming to elliptical co-ordinates, 
\begin{equation}
 D_{rr}  =  n \lambda^3 \cdot \frac{V}{\lambda}  \cdot \lb \frac{\kappa}{V \lambda} \rb^2 \cdot I \lb \frac{a}{\lambda} \rb
\label{e:dipScal}
\end{equation}
where $I$ is the integral with the dimensional constants taken out of the integration: 
\numparts
\begin{eqnarray}
&&I(a/\lambda)=\frac{1}{d}\int \xi^2(\sigma,\tau) d \Omega(\sigma,\tau), \label{e:Integral} \\ 
&& \xi = \vert \Delta \vert \lb \frac{2 \kappa}{V \lambda} \rb^{-1}, \\
&&d \Omega = 2 \pi \, (\sigma^2 - \tau^2) \, d \sigma \, d \tau. 
\end{eqnarray} 
\endnumparts
Here the integration is carried over all $\tau$ and $\sigma$ such that initial position of the tracer, given by (\ref{e:elliptica}) and (\ref{e:ellipticb}), lies further away than the distance $a$ from the swimmer path. 

We now identify each of the factors on the rhs of (\ref{e:dipScal}) .
The first factor corresponds to the number of swimmers within one flight of length $\lambda$ from the tracer. The second factor accounts for the number of statistically independent path segments. The third factor corresponds to a characteristic tracer displacement in an interaction with a single path segment of a swimmer (`collision') at distances $O(\lambda)$. The last factor, i.e. the function $I \lb a/\lambda \rb$, arises from the re-scaled lower integration limit in (\ref{e:DD}). (Convergence of the integral at the upper limit does not pose problems.) In the critical phenomena language it may be thought of as the scaling function accounting for the influence of the microscale $a$ at the mesoscale $\lambda$.

It turns out that as $a/\lambda \to 0$ the integral $I$ converges to $I(0)=4 \pi/3$. 
Hence the powers of $\lambda$ in (\ref{e:dipScal}) cancel out! Thus, for $a \ll \lambda$ we recover the scaling law (\ref{e:scalDrr}) and
the surprising conclusion that stirring due to the curvature of swimmer trajectories  turns out independent of the mean curvature radius:
\begin{equation}
D_{rr} =  \frac{4 \pi}{3} \lb \frac{\kappa}{V} \rb^2 n V.
\label{e:Drr}
\end{equation}

 This happens due to a fortuitous balance between the dimensionality of space and the scaling exponent of the force dipole. We also obtain the exact value for the constant $A=4 \pi/3 \approx 4.2$, which is in a reasonable agreement with the value obtained in \cite{LinChildress11}. Note however, that so far we have implicitly assumed that $\lambda \ll L$, where $L$ is the macroscopic system size. This scale separation condition resolves the paradox introduced in section \ref{s:reorient}: as $\lambda$ grows it will inevitably become comparable to the system size and the scaling (\ref{e:scalDrr}) will break down.
This condition is also likely to be violated in many systems of practical interest, such as small droplets, microfluidic devices and microscopic microbial marine environments.  Therefore, we discuss this case in section \ref{s:Simul}.

In real suspensions of swimmers, tracer mixing occurs due to both stirring mechanisms simultaneously. It is reasonable to assume as a first approximation that their effects add up. Therefore the total diffusion
%\vspace{-3mm}
\begin{eqnarray}
D \approx D_{rr} + D_{entr} + D_{therm},
\label{e:Dsum}
\end{eqnarray}
%\vspace{-3mm}
where $ D_{therm}$ is the contribution due to thermal noise, which depends on the physical properties of the diffusing agents, which may be colloidal particles \cite{Leptos}, large polymer molecules \cite{KimBreuer04}, or non-motile bacteria \cite{Poon13}.

%------------
\section{Statistics of tracer particle displacements \label{s:Statistics}}

Experiments on passive tracers in suspensions of eukaryotic swimmers {\it C. reinhardtii} \cite{Leptos} found that the distribution of tracer particle displacements has an unusual form: it consists of a Gaussian core and robust heavy tails. The presence of heavy tails indicates that the system is far from equilibrium and is a warning that the appealing notion of the effective `bacterial thermal bath' \cite{Wu00,Dombrowski04} should be used with caution. 

The heavy tails found in \cite{Leptos} appeared to be exponential and the time-dependent displacement distribution, rescaled with its dispersion, was found to approach a self-similar form. Lin, Thiffeault and Childress \cite{LinChildress11} studied the distributions numerically for a suspension of a particular type of model swimmer, a squirmer, moving with random reorientations. After a short transient, they observed a diffusive regime and also found a self-similar tracer distribution with exponential tails. Further, they related the exponential form of the tails to a particular feature of squirmers, the presence of stagnation points at their surface. These exponential tails, however, became pronounced only after very long integration times, unlike the tails observed by Leptos \etal \cite{Leptos}. Thus, it remained unclear whether these findings were related. 

In this section we revisit the problem of tracer distributions in suspensions of dipole swimmers with random reorientations using our analytical technique. We consider only the far-fields of swimmers and neglect the entrainment. 

The probability distribution of dimensionless tracer displacements due to a single straight swimmer path
\begin{eqnarray}
P(\xi')= (1/Q) \int \delta(\xi'-\xi(\sigma,\tau)) d \Omega(\sigma,\tau),
\label{e:Pxi}
\end{eqnarray}
where the value of the constant $Q$ is set by the normalisation condition. The characteristic function for the uncorrelated isotropic random walk with steps $\bmxi$ having the distribution $P(\xi)$, where $\xi=\|\bmxi\|$,
\begin{eqnarray}
g(k) = \langle e^{i \bk \cdot \bmxi} \rangle = \int_0^\infty \frac{\sin(k \xi)}{k \xi} P(\xi) d \xi.
\label{e:g}
\end{eqnarray}
The behavior of $P(\xi)$ for large $\xi$ is reflected in the asymptotic behaviour of the characteristic function at small $k$. One could study it analytically using the expression for $g(k)$ obtained after substituting (\ref{e:Pxi}) in (\ref{e:g}). However, the result can be obtained more easily by using the Mellin transform of the distribution of $\xi^2$, $P_{\xi^2}(\xi^2) d(\xi^2)= P(\xi) d \xi$:
\begin{eqnarray}
M(s)=\int_0^\infty \xi^{2(s-1)} P_{\xi^2}(\xi^2) d(\xi^2) = \int_0^\infty \xi^{2(s-1)} P(\xi) d \xi.
\end{eqnarray}
Substituting the expression (\ref{e:Pxi}) we obtain
\begin{eqnarray}
M(s) \propto \int \xi^{2(s-1)}(\sigma,\tau) d \Omega(\sigma,\tau).
\end{eqnarray}
The integrand of the latter expression is a rational function amenable to analytical treatment. It can be readily checked that it has a simple focus at $s_*=5/2$ and this focus is the rightmost singularity of $M(s)$ in the complex plane.
According to the theory of the Mellin transform \cite{Flajolet}, $P_{\xi^2} \sim \xi^{-2s_*}=\xi^{-5}$ for large $\xi$. Hence  $P(\xi) \sim \xi^{-4}$.

This result can be understood using a simple physical argument. We observe that the largest tracer displacements occur when the starting or ending points of the straight swimmer path happen to be in the vicinity of the tracer. For a uniform distribution of swimmers the probability of an endpoint at a distance $r$ from a tracer is proportional to the spherical volume of radius $r$.  As $\xi \sim 1/r$, the probability distribution of tracer displacements for large $\xi$
\begin{eqnarray}
P(\xi) d \xi \propto d(r^3) \propto d ( \xi^{-3} ) \propto \xi^{-4} d \xi.
\end{eqnarray}

The finite variance of the distribution guarantees by virtue of the Central Limit Theorem that in the limit of large number of swimmers, the distribution of the tracer displacements approaches a Gaussian. Note that at the same time the Central Limit Theorem is violated for fluid velocity fluctuations \cite{Rushkin,Zaid11}. Also, although the distribution of tracer displacements approaches a Gaussian in the central region, the heavy power-law tails of the distribution survive. 

Indeed, by a standard argument of random walk theory, the characteristic function after $N$ independent steps $g_N(k)=g(k)^N$. In our case $N$ equals the number of independent straight swimmer path segments $N=n \lambda^3 V t/\lambda = n \lambda^2 V t$. It is a well-known fact, known as a Tauberian theorem in the mathematical literature \cite{Bazant}, that distributions with a finite variance $\sigma_0^2=\langle \xi^2 \rangle$ having power-law tails $\sim C \xi^{-4}$ have a characteristic function $g(k)= 1 - \half k^2 \sigma_0^2 - \const \cdot C \vert k \vert^3 + O(k^{-4})$. Here the second term corresponds to the Gaussian core of the distribution function, while the third term, which is singular because it is a function of $\vert k \vert$, corresponds to the power-law tails of the distribution. It follows immediately that the power-law tails survive outside the (broadening) Gaussian core for large $N$.

Therefore we conclude that the random reorientations mechanism alone does not produce self-similar tracer statistics. This conclusion contrasts with the claims of Leptos \etal \cite{Leptos} and conclusions of Lin \etal \cite{LinChildress11}. However, the experimental data \cite{Leptos} was obtained for relatively short times, much shorter than the mean reorientation time, while the results \cite{LinChildress11} were obtained for extremely long times and were related to a rather specific feature of the swimmer model considered. Thus, there is no direct contradiction with our conclusions. 

%------------
\section{Numerical simulations of stirring in confined systems \label{s:Numerics}}

\subsection{Simulations description}

In order to validate our theoretical predictions and estimate the effects of approximations, we performed numerical simulations using a specific model of dipole swimmers, spherical squirmers \cite{Lighthill_squirmer,Blake_squirmer}. Similar simulations were carried out in \cite{LinChildress11}. 

A squirmer is a sphere moving in viscous fluid with (tangential) velocity prescribed at its boundary. The flow field created by a squirmer is given analytically as an expansion in spherical harmonics. Squirmers have been used to describe the motion of ciliates propelled by metachronal waves of cilia on their surface. Following Ishikawa \etal \cite{Ishikawa_squirmer} and Lin \etal \cite{LinChildress11} we choose a specific squirmer model with non-zero first two modes. While the strength of the first squirmer mode sets the swimming velocity $V$, the strength of the second mode sets the (dimensionless) swimmer dipole strength $\beta$. While in the far-field a squirmer velocity field is dominated by the dipole term, in the near field it is non-singular due to a regularising quadrupole term. 

In our simulations the time and space units were chosen such that the swimmers' size and speed $a=V=1$. Clearly, in this system the reorientation distance $\lambda$ is also the reorientation time. The system size in our simulations is controlled by the size of the cubic computational domain $L$. 
Interactions between swimmers are neglected and the fluid velocity field is obtained as a superposition of the fields of individual squirmers. The concentration of squirmers is kept low, $\Phi \sim 10^{-6}-10^{-5}$, so that collisions between them are unlikely. When a swimmer leaves the computational box it is injected with the same orientation at a diametrically opposite location, so that the total number of swimmers remains constant. However, this process introduces a discontinuity in the swimmer velocity field of the order $O((a/L)^2)$. We control this effect by making $L$ sufficiently large. The time integration of tracer trajectories was performed using a semi-implicit second order Euler method with a varying time step. The simulations were run for the dimensionless time $\Delta T = 10^4$ and for 300 independent realisations.

\subsection{Simulations results \label{s:Simul}}

\begin{figure}
\includegraphics[width=0.89\columnwidth]{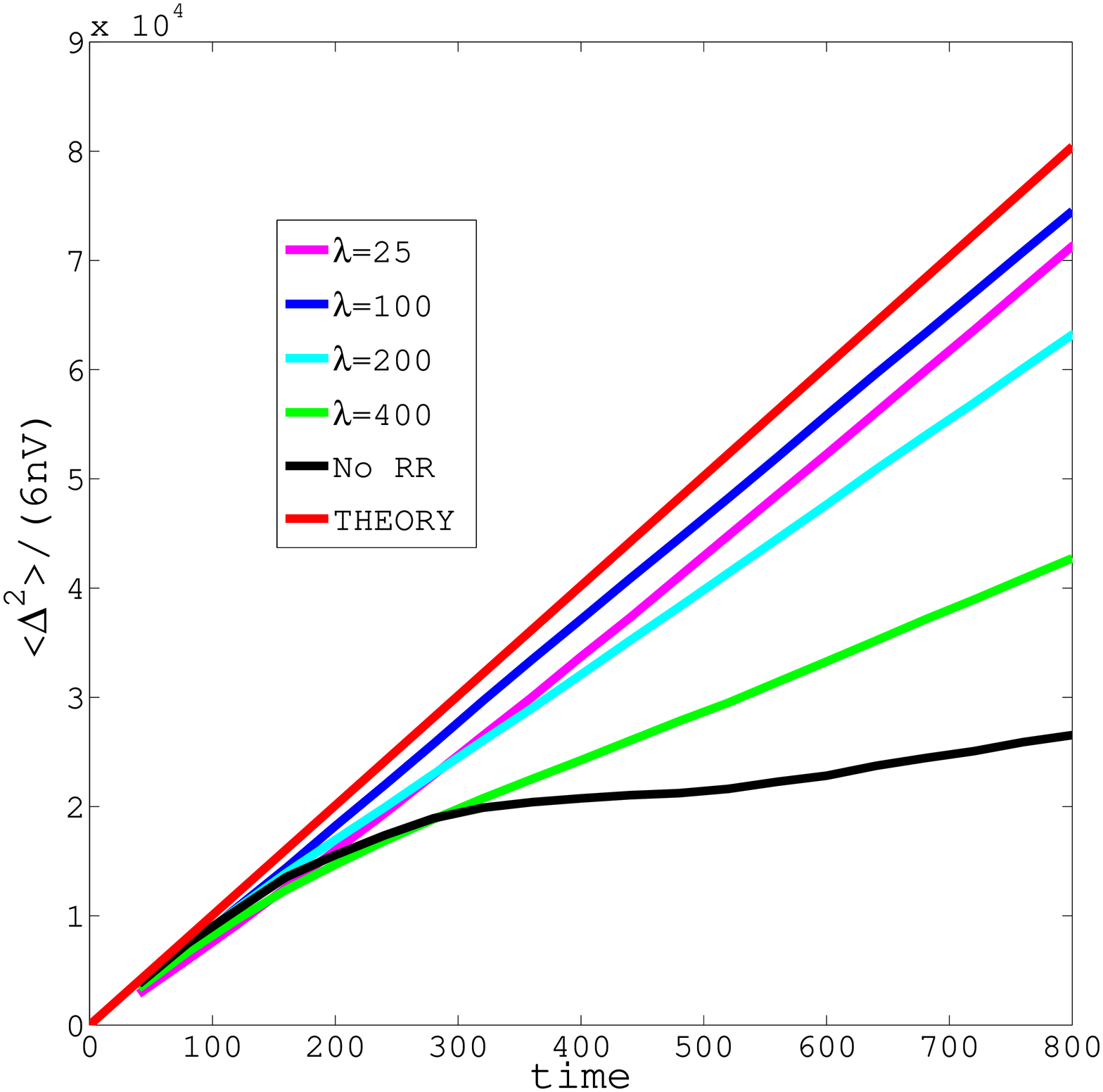}
\includegraphics[width=0.89\columnwidth]{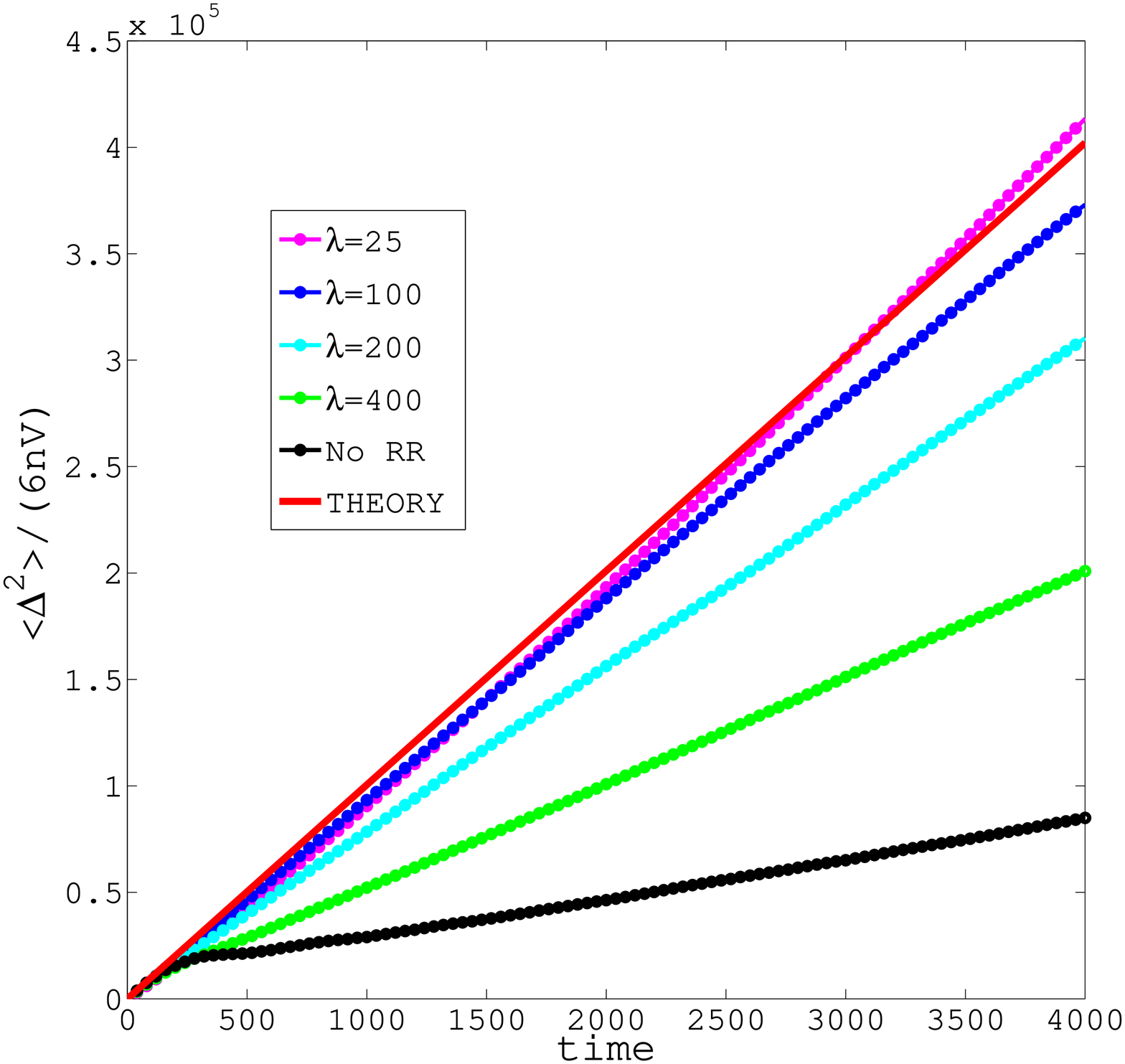}
\includegraphics[width=0.89\columnwidth]{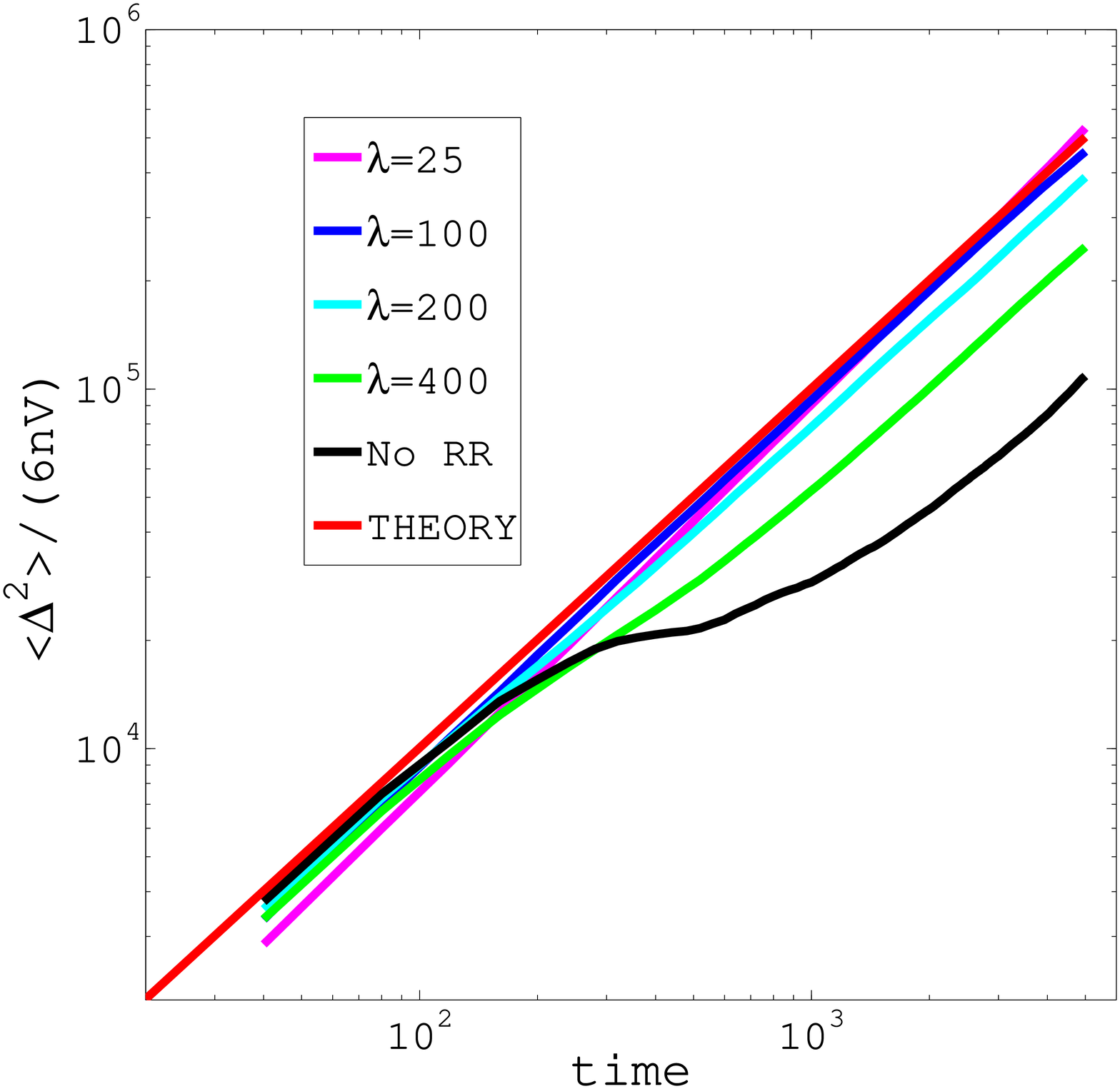} 
\caption{Mean squared displacement of a tracer as a function of time. (a) Short times. (b) Longer times. (c) In double logarithmic coordinates. The lines computed for varying $\lambda$ fill the envelope between the straight red line corresponding to $D_{rr}$ calculated according to formula (\ref{e:Drr}) and the black line computed for swimmers with infinite straight trajectories. The initial transient convexity of these lines, most pronounced for swimmers moving with no reorientations, is a manifestation of the tendency of tracer trajectories to form loops. In these simulations the dimensionless dipole strength of swimmers $\beta=2$ and the computational box size $L=500$. \label{f:MSDt}}
\end{figure}

\begin{figure}
\includegraphics[width=0.89\columnwidth]{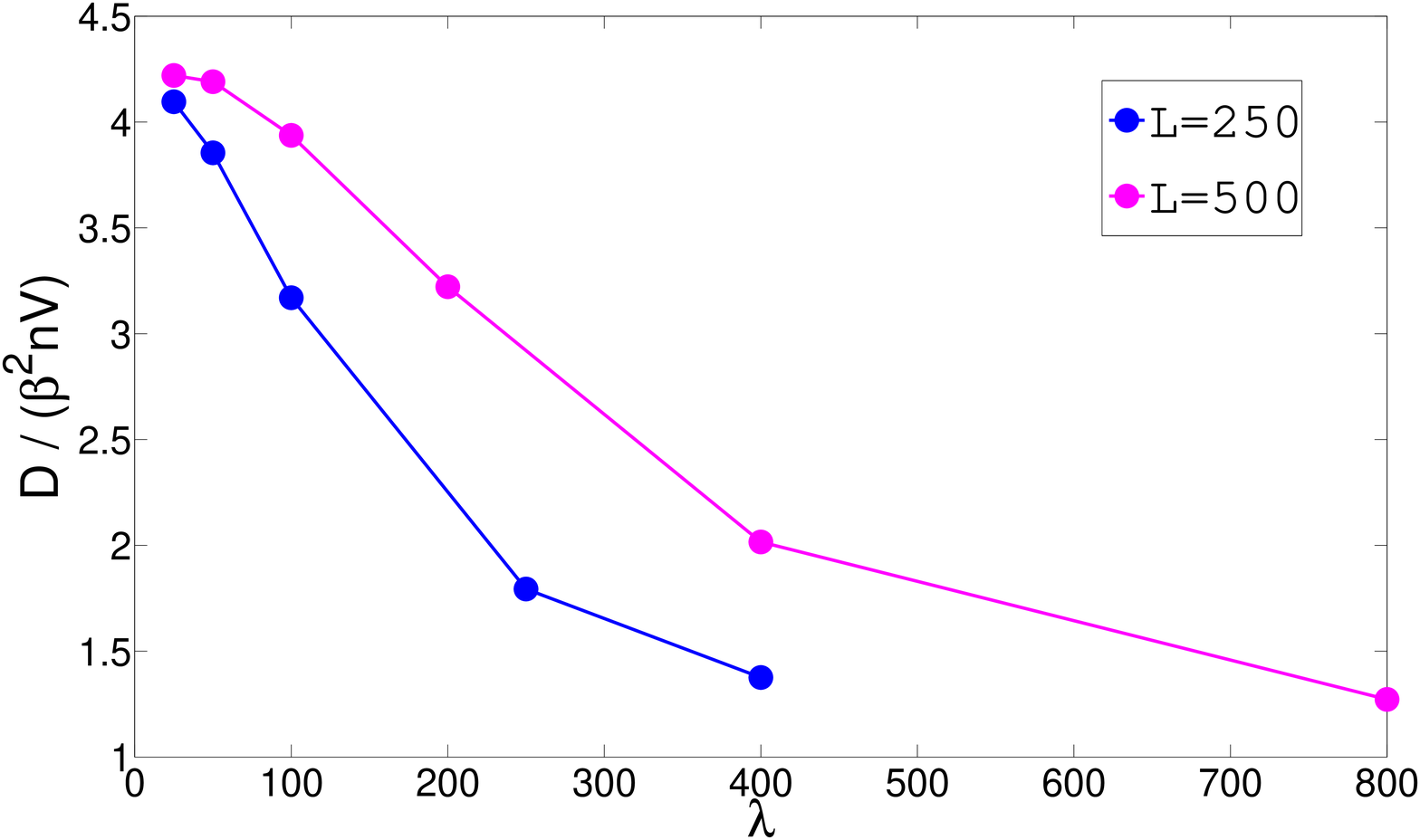}
\includegraphics[width=0.89\columnwidth]{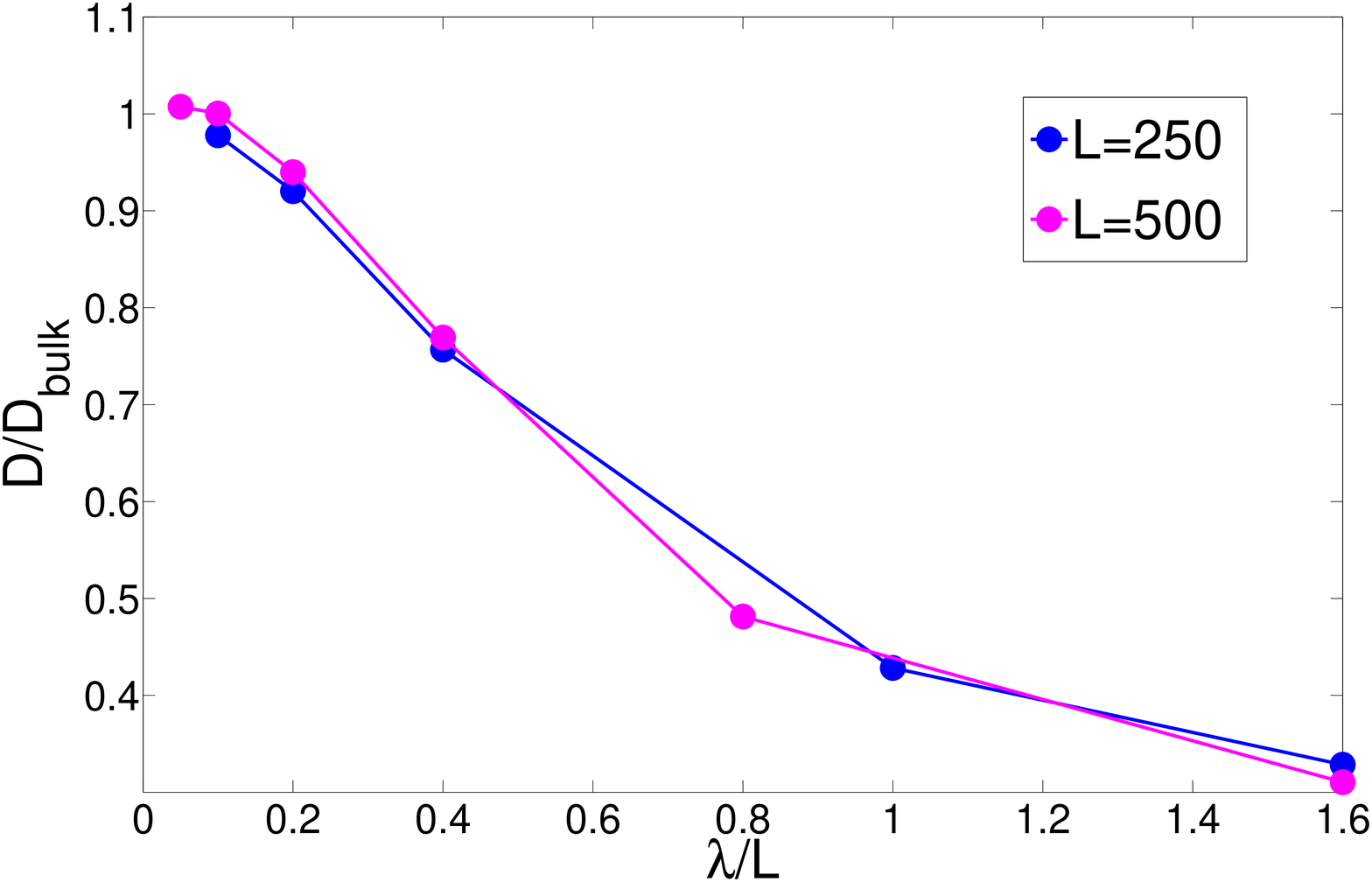}
\caption{Tracer diffusion for systems of finite size computed in numerical simulations of a suspensions of squirmers. The two curves are computed for varying $\lambda$ and the computational box sizes $L=250$ and $500$. The curves collapse onto a single scaling curve when plotted as functions of $\lambda/L$. Here the tracer diffusion coefficient for the bulk $D_{bulk}$ is calculated according to expression (\ref{e:Drr}). \label{f:Dcollapse}
}
\end{figure}

Firstly, our simulations show that after an initial transient the tracer particle motion becomes diffusive and $\langle \Delta^2 \rangle \sim t^2$, see figure \ref{f:MSDt}. The duration of the transient depends on the random reorientation length, becoming longest for swimmers with infinite straight trajectories (having no reorientations), see figure \ref{f:MSDt}(a). During the transient the function $\langle \Delta^2 \rangle(t)$ is initially convex and could be fitted by $\langle \Delta^2 \rangle \sim t^\alpha$ with $\alpha<1$. 
We would like to point out that in dilute swimmer suspensions this behaviour is a manifestation of the loop-like tracer trajectories described in section \ref{s:Intro}. At very early times the tracer displacement is proportional to the local flow velocity and time correlation effects are not felt. Hence the resulting tracer motion is diffusive \cite{Zaid11}. As time progresses, the loop effect comes into action and halts the effective growth of $\langle \Delta^2 \rangle$ making it a convex function of time. Note that for swimmers with infinite straight trajectories, the characteristic time for a tracer to turnaround during its interaction with a particular swimmer depends on the distance between the tracer and the trajectory. Therefore for an ensemble of swimmers the turnaround time has a broad distribution rather than a single well-defined value. Consequently, for such swimmers the transition to the long time diffusive regime is rather protracted. This is clearly seen on figure \ref{f:MSDt}.

By contrast, when swimmers have a finite reorientation length and, equivalently, reorientation time,  the memory of the system does not extend beyond it. Therefore the tracer motion becomes strictly diffusive at times larger than the reorientation time.

The theory summarised in section \ref{s:Stirring} predicts that the diffusion due to random reorientations is given by expression (\ref{e:Drr}), which, rather remarkably, is independent of the reorientation length. The straight line corresponding to the diffusion coefficient calculated according to (\ref{e:Drr}) for $\beta=2$ is drawn in figure \ref{f:MSDt} in red. The lines of other colour were obtained in numerical simulations with the computational box of fixed length $L=500$ and correspond to different values of the random reorientation length $\lambda$. For relatively short $\lambda$ the computed lines closely approach the theoretical line, while for $\lambda \to \infty$ they approach the no-reorientation line, filling the envelope between the theoretical and the no-reorientations lines for intermediate values of $\lambda$. 

\begin{figure}
\includegraphics[width=0.89\columnwidth]{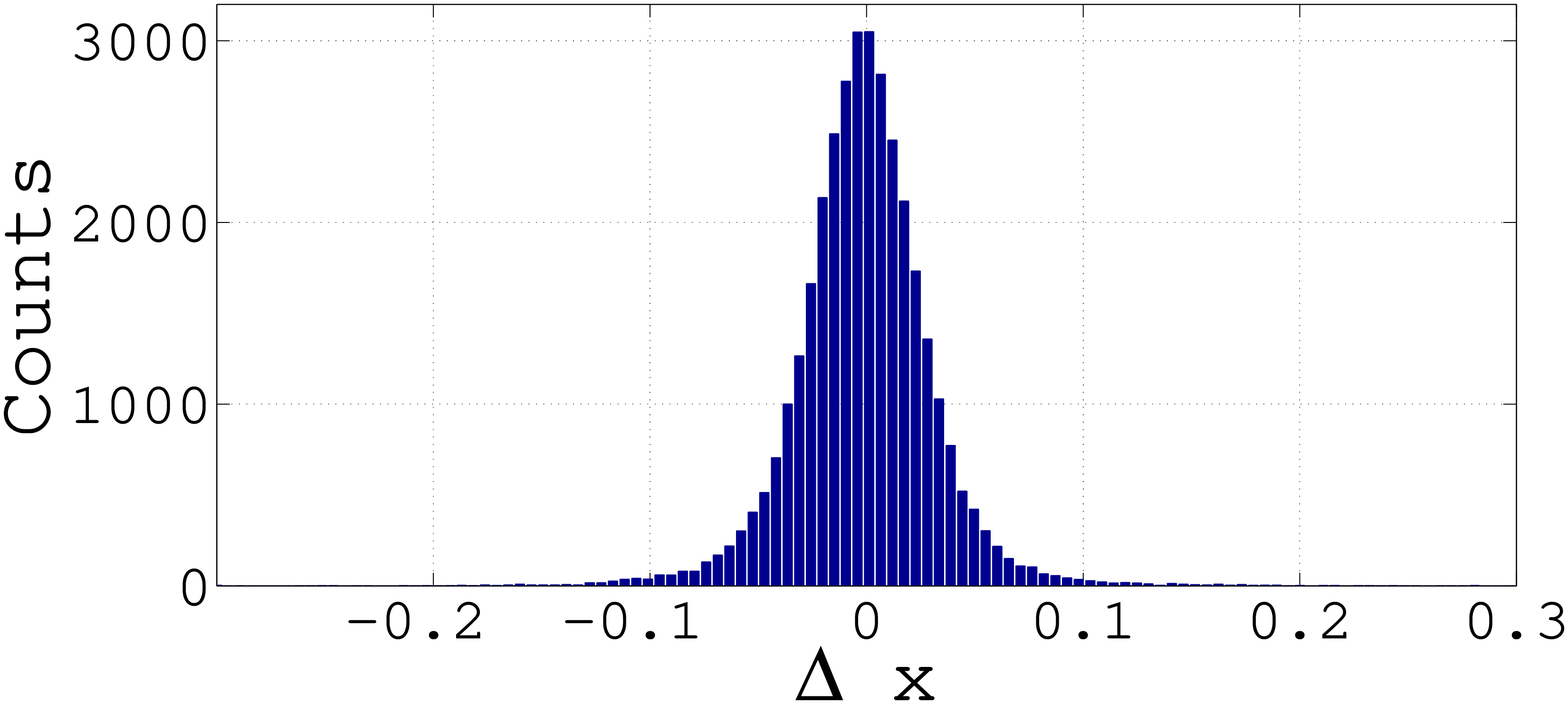}
\includegraphics[width=0.89\columnwidth]{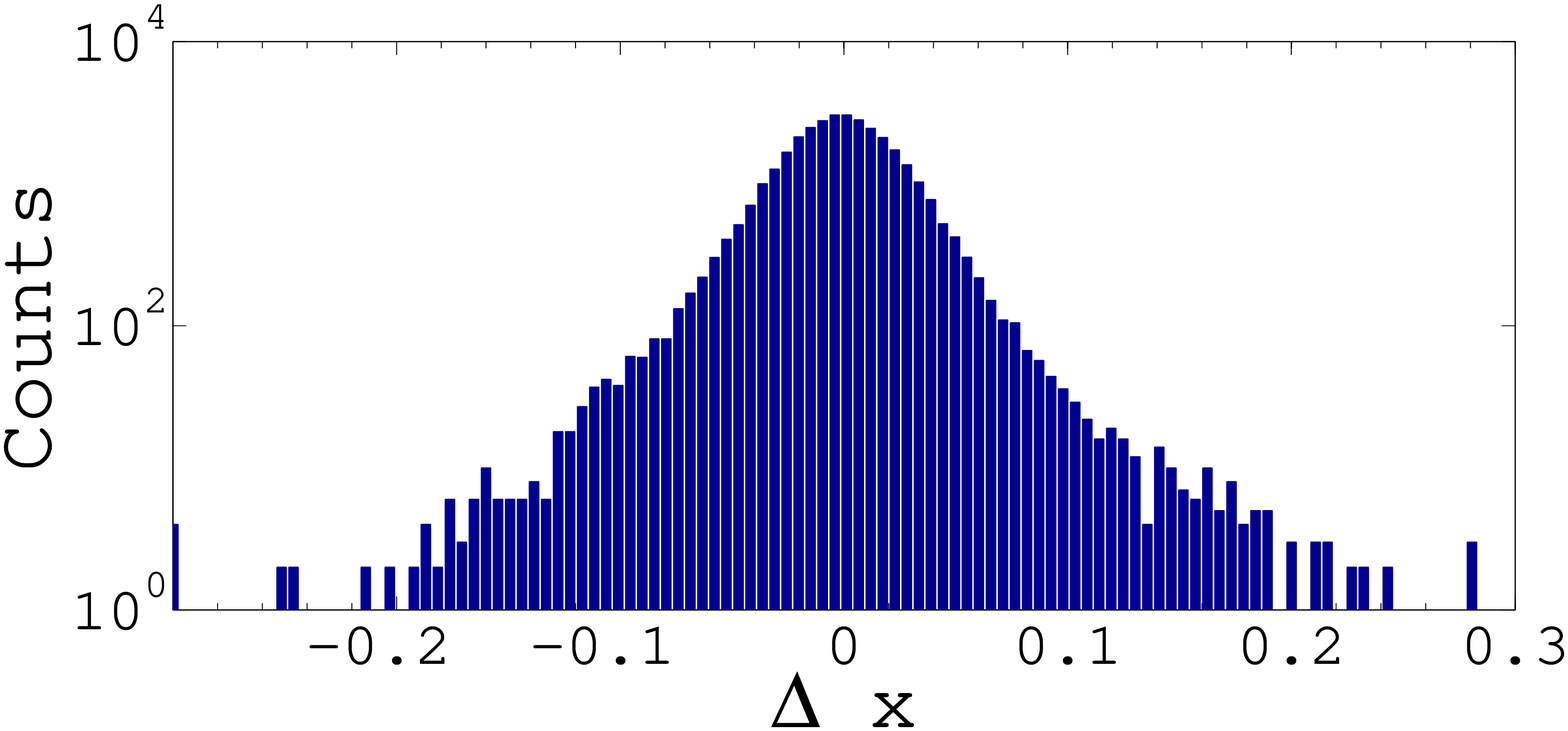}
\includegraphics[width=0.89\columnwidth]{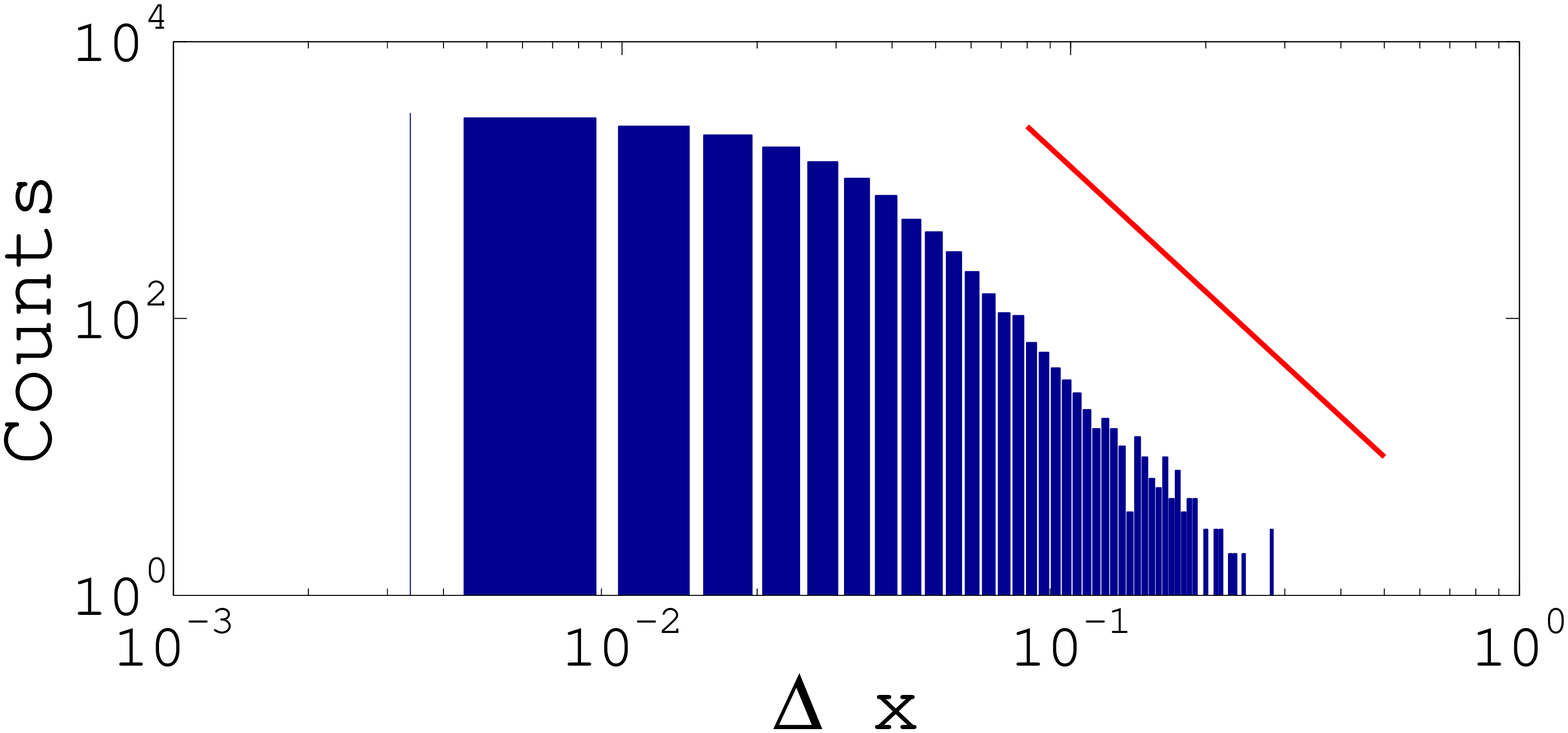}
\caption{Histogram of tracer displacements (along the x-axis) at time $t=120$. (a) The distribution is visibly close to a Gaussian. (b) Heavy tails of the distribution become visible on the semi-log plot. (c) The power-law tails at large deviations correspond to the power law $p(\Delta) \sim \Delta^{-3}$ shown on the plot by the straight red line. The parameters of the simulation: the reorientation length $\lambda=100$, the swimmer dipole strength $\beta=2$, the number of swimmers $N=400$ and the computational box size $L=1000$.}
\label{f:Histo}
\end{figure}

Several conclusions can be drawn from these observations. Firstly, the tracer diffusion due to random reorientations seems to depend on $\lambda$. However, we will further show that this dependence arises solely due to a finite system size. 

Secondly, the effective tracer diffusion is dominated by random reorientations (equivalently, curvature of swimmers' trajectories) for 3D dipole swimmers. In the current setting, however, this conclusion could have been expected as, according to the expression (\ref{e:tau_c}), the time between the entrainment events $\tau_c \sim 10^4$ and is of the same order as the simulation time. Therefore, the tracer is relatively rarely entrained by the swimmers. In order to resolve the question of the relative magnitude of $D_{entr}$ and $D_{rr}$ much longer simulations would be needed. Note, however, that the present numerical setting reflects the situation commonly encountered in experiments.

Finally, an excellent agreement of the computed value of $D$ with the prediction of the expression (\ref{e:Drr}), validates the theoretical model \cite{PushPRL}. In particular, it confirms the quadratic dependence of $D_{rr}$ on the swimmer dipole strength $\kappa$. 
%{\color{blue} This conclusions agrees with the previous numerical results \cite{LinChildress11} but not the more recent simulations \cite{Morozov13}. }

%\subsection{The effect of confinement \label{s:Finite}}  

In a variety of physically and biologically important situations the reorientation length $\lambda$ becomes comparable to the system size $L$ \cite{Stoker_microfluidics,Poon_agar,MarineEcosystems07,MarineEcosystems85}. This condition leads to a breakdown of the formula (\ref{e:Drr}) for $D_{rr}$. However,  (\ref{e:Drr}) can be easily generalised. Indeed, by a dimensional theory argument, for $a \ll \lambda \sim L$ the only relevant large length scales are $\lambda$ and $L$. Hence, $D_{rr}$ will assume the form
\begin{eqnarray}
D_{rr} =  \frac{4 \pi}{3} \lb \frac{\kappa}{V} \rb^2 n V g \lb \lambda/L \rb,
\label{e:DD_finite}
\end{eqnarray}
where $g(x) \to 1$ for $x \to 0$ and $g(x) \to 0$ for $x \to \infty$.

Our numerical simulations, indeed, corroborate this prediction. Figure \ref{f:Dcollapse} shows that the values of $D$ computed for two different box sizes $L=250$ and $L=500$ collapse on the same curve when potted as functions of $\lambda/L$. It should be clear that the scaling function $g(x)$ in this setting is universal, i.e. independent of the concentration, the swimming speed and the dipole strength of swimmers.
 
Figure \ref{f:Histo} presents the distribution of passive tracer displacements obtained in simulations at time $\Delta t = 120$. Notice the Gaussian character of the central region of the distribution and formation of clearly visible power-law tails for large tracer displacements, in agreement with the general theoretical predictions of section \ref{s:Statistics}. The power-law tails appear somewhat broader than predicted by the theory. Very large tracer displacements are likely to be affected by the entrainment mechanism as they occur when a swimmer passes close to the tracer and the tracer velocity can not be neglected in comparison with that of the swimmer.

%----------
\section{Discussion \label{s:Discussion}}

Our previous theoretical work \cite{PushPRL} was contingent on a number of assumptions, such as the statistical independence of contributions from consecutive swimmer runs and ignoring the near-field effects in calculating the tracer displacement using (6). The quantitative agreement of our current simulations with the theory confirms that these assumptions are justified. A good agreement of the mixing rates predicted by (\ref{e:Drr}) with experimental measurements had been established previously \cite{PushPRL}.

We also confirmed that $D_{rr}$ gives the dominant contribution to the total tracer diffusion coefficient $D$, see (\ref{e:Dsum}). However our current simulations are not extensive enough to settle the question about the relative magnitude of $D_{rr}$ and $D_{entr}$ quantitatively, because the entrainment events for an individual tracer are relatively rare. This situation, however, is typical for experimental settings.  Therefore we expect that in such circumstances the enhanced diffusivity of passive particles does not depend on particular features of the microorganisms' locomotion and on their trajectories' persistence length, unless the latter becomes comparable with the system size $L$. For longer times and swimmers having a weak dipole strength $D_{entr}$ may turn out important. Then the enhanced passive particle diffusivity will depend on the details of swimmers' locomotion \cite{Pushkin13a}.

Our analysis presumes that the swimmers have a well-defined run length $\lambda$. In reality many bacteria, e.g. {\it E. coli}, execute run-and-tumble motion with an exponential distribution of run lengths \cite{Berg72}. This factor has been accounted for in some theoretical studies \cite{Koch2012}. However, in the present context it leads to no new effects. Because $D_{rr}$ does not depend on $\lambda$, it will not change after averaging over the distribution of run lengths. Other conclusions do not change significantly either.

Our most important observation is the effective reduction of $D_{rr}$ due to finite system size effects. This factor should be of particular interest to ecological and technological applications, as examples of microenvironments populated by swimmers are numerous and include small droplets, microfluidic devices and microzones with an abundant supply of nutrients in marine environments. Confinement may influence the stirring in a variety of ways that include the effects of the accumulation of swimmers close to the boundaries \cite{Shum2010} and alteration of the velocity fields of swimmers due to the (slip or no-slip) boundary conditions. The latter effect becomes important when the system size L becomes comparable to the swimmer size a. Here, however, we emphasize that the confinement will be felt much earlier, when the system size becomes comparable to the reorientation length $\lambda$, i.e. for $a << \lambda ~ L$. It will also be felt in systems with no external mechanical boundaries \cite{MarineEcosystems07,MarineEcosystems85}. This is the effect considered in our work.

A decrease of the effective tracer diffusion for large $\lambda$ was also observed in recent simulations reported in \cite{Morozov13}.  The authors considered stirring in a suspension of run-and-tumble swimmers having a purely dipolar velocity field regularised close to the swimmer by the rule that the tracer velocity is put to zero if the tracer is closer than approximately one swimmer radius from the centre of the swimmer. 
The authors did not attribute the observed decrease of stirring to the finite system size effect, but rather hypothesised that such sensitivity of stirring to the swimmer parameters may be an intrinsic feature of biogenic mixing. Our results do not uphold this conclusion. 
%(At the same time, the authors confirmed that thermal fluctuations give an additive contribution to the total mixing rate, as assumed in expression (\ref{e:Dsum}) \cite{PushPRL}. )

We find that the ratio of the effective diffusion coefficient in confined environments to its bulk value $D/D_{bulk}$ is given (at constant swimmer concentrations) by a universal function $g(\lambda/L)$ and shows a decrease by the factor of two when $\lambda \simeq 0.8 L$. It is important to remember, however, that in ecological microenvironments sustained by a localised supply of nutrients, the local concentrations of bacteria are often orders of magnitude higher than in the surrounding bulk \cite{MarineEcosystems07} and the overall mixing is enhanced.

\section*{References}

\bibliographystyle{unsrt}
\bibliography{Reorientations_bib}

\end{document}